# Enhanced Ferromagnetism in Cylindrically Confined MnAs Nanocrystals Embedded in Wurtzite GaAs Nanowire shells

*Anna Kaleta[1]\*, Slawomir Kret[1]\*, Katarzyna Gas[1], Boguslawa Kurowska[1], Serhii B. Kryvyi[1], Bogdan Rutkowski[2], Nevill Gonzalez Szwacki[3], Maciej Sawicki[1] and Janusz Sadowski[1,4]\**

[1] Institute of Physics, Polish Academy of Sciences, Aleja Lotnikow 32/46, PL-02-668 Warszawa, Poland

[2] Faculty of Metals Engineering and Industrial Computer Science, AGH University of Science and Technology, Aleja A. Mickiewicza 30, 30-059 Kraków, Poland.

[3] Institute of Theoretical Physics, Faculty of Physics, University of Warsaw, Pasteura 5, 02-093 Warszawa, Poland.

[4] Department of Physics and Electrical Engineering, Linnaeus University, SE-391 82 Kalmar, Sweden

Nearly a 30% increase in the ferromagnetic phase transition temperature has been achieved in strained MnAs nanocrystals embedded in a wurtzite GaAs matrix. Wurtzite GaAs exerts tensile stress on hexagonal MnAs nanocrystals, preventing a hexagonal to orthorhombic structural phase transition, which in bulk MnAs is combined with the magnetic one. This effect results in a remarkable shift of the magneto-structural phase transition temperature from 313 K in the bulk MnAs to above 400 K in the tensely strained MnAs nanocrystals. This finding is corroborated by the state of the art transmission electron microscopy, sensitive magnetometry, and the first-principles calculations. The effect relies on defining a nanotube geometry of molecular beam epitaxy grown core−multishell wurtzite (Ga,In)As/(Ga,Al)As/(Ga,Mn)As/GaAs nanowires, where the MnAs nanocrystals are formed during the thermal-treatment-induced phase separation of wurtzite (Ga,Mn)As into the GaAs−MnAs granular system. Such a unique combination of two types of hexagonal lattices provides a possibility of attaining quasi-hydrostatic tensile strain in MnAs (impossible otherwise), leading to the substantial ferromagnetic phase transition temperature increase in this compound.

**INTRODUCTION**

Various forms of strain manipulation including thermal effects, phase change, dislocations or periodic lattice strain engineering have been shown to improve and modify physical properties of functional materials.[1,2] Importantly, the unprecedented progress in a spatial resolution of modern characterization techniques as well as in the first principles methods efficiency have simultaneously allowed a quantitative understanding of the mechanisms responsible for these modifications.[2,3] The hybrid structures comprising magnetic nanocrystals (NCs) embedded in semiconducting matrices, constitute a range of materials with magnetic properties strongly dependent on the various forms of strain.[4,5] Such hybrid systems can offer functionalities similar to those expected from dilute ferromagnetic semiconductors (DFS),[6] but they enable overcoming the low ferromagnetic transition temperature ($T_C$) limit of DFS, known not to exceed 200 K.[7] Therefore, since the very beginning of the research activity devoted to DFS, phase segregated, hybrid materials consisting of magnetic NCs embedded in a semiconductor host had been investigated.[8] Such a composite material can be fabricated the easy way, by a medium-to-high temperature (400 – 600 °C) post-growth annealing of (Ga,Mn)As, the canonical DFS, resulting in the GaAs:MnAs hybrid system.[9,10,11,12,13,14,15] Since the equilibrium solubility of Mn in the GaAs host lattice is low (below about 0.2%),[16] the (Ga,Mn)As ternary compound can only be obtained by highly nonequilibrium methods such as low temperature molecular beam epitaxy (MBE),[7] or ion implantation.[17,18,19] Due to the coexistence of semiconducting and ferromagnetic properties, (Ga,Mn)As exhibits a wealth of interesting magnetotransport properties. However, all the prospecting features of this class of materials can be harnessed only at the cryogenic temperatures. This obstacle hampers employment of (Ga,Mn)As in the application-ready devices and fuels ever so growing interest in the hybrid systems. Despite their inferior electrical properties (e.g., conductivity) to those of (Ga,Mn)As (or other DFSs), they offer much higher, technologically viable $T_C$, which is usually considerably higher than the room-temperature (e.g., MnSb microcrystals embedded in GaSb,[20] or ε-$Fe_3N$ NCs embedded in GaN,[21] both with $T_C$ in the vicinity of 300 °C).

Here, we investigate the formation of a hybrid nanoscale system consisting of MnAs NCs embedded in the wurtzite (WZ) GaAs. MnAs is a ferromagnetic metal assuming it is in its native (bulk) low-temperature phase, the hexagonal NiAs-type crystalline structure. The ferromagnetic-paramagnetic phase transition at about 313 K (40 °C) is associated with the structural one from the hexagonal, NiAs-type (α-MnAs), to the orthorhombic MnP-type (β-MnAs) phase.[22,23,24] This structural transition is combined with the 2% volume contraction mainly due to the decrease of a basal plane lattice parameter $a$.[25] All of the previous literature reports on GaAs:MnAs hybrids concerned planar samples, with MnAs nanoinclusions embedded in the zinc-blende GaAs thin films. The moderate influence of strain on the magnetic properties of MnAs both in the form of epilayers deposited on, and NCs embedded in the zinc-blende GaAs had already been noticed, but resulted in a moderate increase of $T_C$.[26,27,28,29]

The significance of our experimental findings is 2-fold. First, we show that the Curie temperature in a ferromagnetic metal MnAs can be substantially increased from the textbook 313 K to above 400 K by the adequate nanostructure engineering. Namely, by employing the core-shell nanowire geometry, we force (Ga,Mn)As shell to undertake the *wurtzite* structure, which then is thermally transformed to the phase separated system containing tensely strained MnAs NCs in the *wurtzite* GaAs surrounding. Second, we evidence the advantage of nanoscale structures (nanowires) with crystallographic structures differing from that of the same materials in their pristine 3D (and 2D) forms; enabling the appropriate strain engineering in hybrid (granular) systems based on such nanowires. Our findings are evidenced by the combined high resolution transmission electron microscopy, sensitive magnetometry investigations and *ab initio* calculations.

**RESULTS AND DISCUSSION**

Core-multishell $Ga_{0.92}In_{0.08}As$-$Ga_{0.60}Al_{0.40}As$- $Ga_{0.94}Mn_{0.06}As$- GaAs NWs have been grown by MBE on GaAs(111)B substrates using gold nanodroplet-assisted vapor-liquid-solid growth mode for WZ $Ga_{0.92}In_{0.08}As$ cores and epitaxial growth conditions for $Ga_{0.60}Al_{0.40}As$, $Ga_{0.94}Mn_{0.06}As$ and GaAs radial shells with a special attention paid to the growth of (Ga,Mn)As in this unique geometry.[30] Typical NWs with the surface density of ~$10^9$ cm$^{-2}$ are shown in the scanning electron microscopy (SEM) image in Figure 1a and detailed in the schematic drawings in the panels b-d in Figure 1. The intended chemical structure of NWs is revealed in the scanning transmission electron microscopy (STEM) images, which are embedded in the panels b and c. In particular, we notice that all of the subsequent shells are in a perfect epitaxial relation with sharp and coherent interfaces, i.e., all the shells are tensely strained to the NW core, while inheriting the WZ core structure.

A part of these as-grown NWs (Figure 1a,b,c) has been subjected to the 20 min postgrowth annealing at 450 ºC, resulting in the (Ga,Mn)As NWs shells transformation into the WZ-GaAs matrix with phase separated ensembles of MnAs NCs, as shown in Figure 1d. The as–grown (nonannealed) NWs were thoroughly analyzed previously.[31]

**Structural investigations.** The structure of the NWs is investigated using transmission electron microscopy (TEM) methods. We examine both the entire NWs separated from the substrate as well as their parts in the form of thin slices cut at various angles by the focused ion beam (FIB) technique. The former enables skipping time-consuming preparation procedures and avoids degradation of the sample material during thinning. The latter gives unambiguous insight into the details of this complex core-shell architecture. In this way we eliminate the features originating from the entire object illuminated by the electron beam and projected onto a 2D image plane.

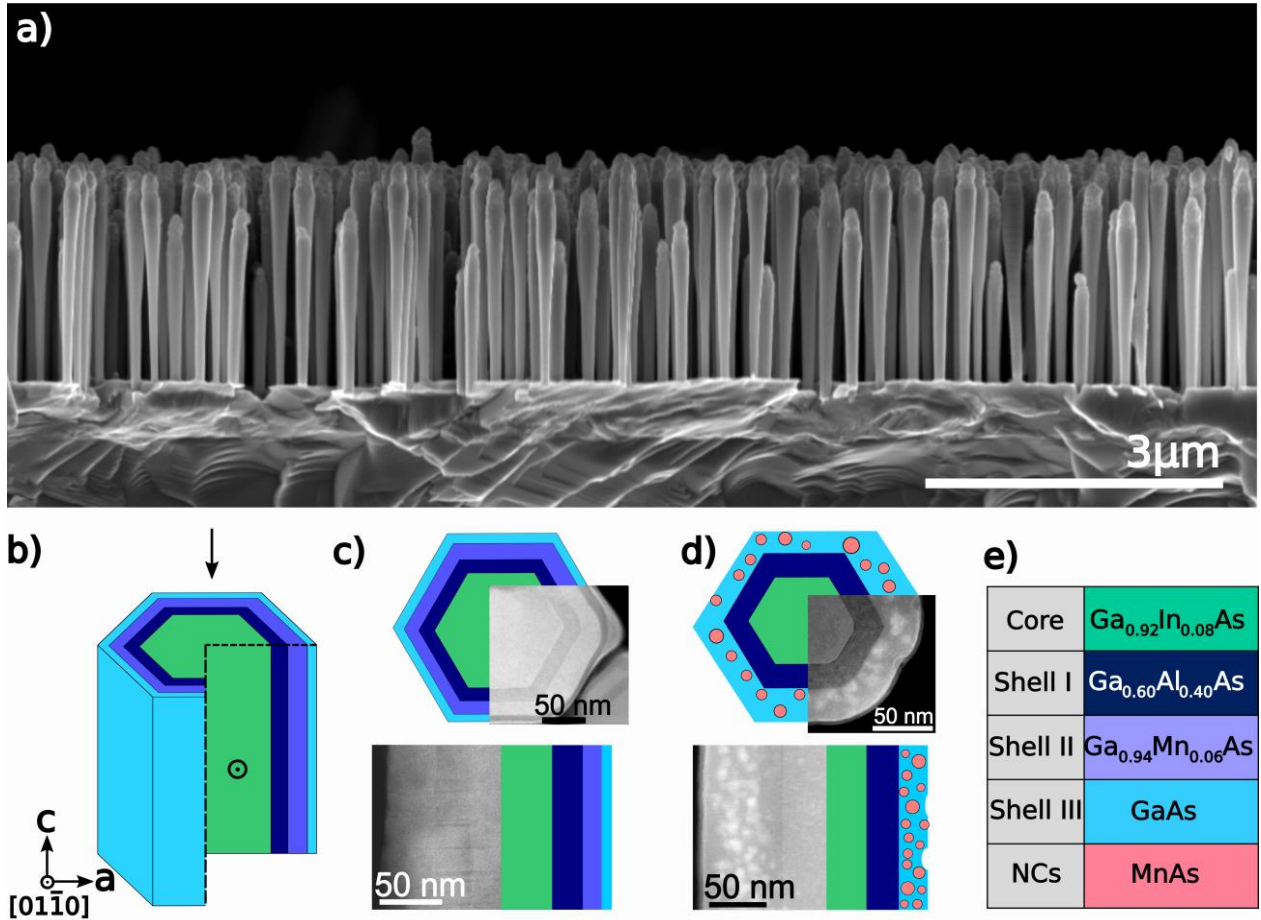

Figure 1. (a) Scanning electron microscopy image of the as-grown NWs. (b) General layout of the as-grown NWs. (c-d) Schematics and scanning transmission electron microscopy visualizations of a sequence of cross-sections seen perpendicularly and parallel to the growth axis of (c) as-grown NWs and (d) NWs annealed at 450 ºC. (e) Table describing chemical composition of subsequent building-blocks of the investigated NWs.

Results of the STEM analysis of the exemplary annealed NWs are presented in Figure 2. For these measurements the NWs are transferred onto a TEM grid. As documented in the panel (a), a typical NW is 2 μm long and its diameter varies from around 70 nm at the bottom part (close to the substrate) to around 250 nm near the top. This diameter variation originates from a shadowing effect occurring during the MBE shell growth.

The images shown in Figure 2 have been acquired in the STEM mode using a high angle annular dark field (HAADF) detector which is sensitive to the average atomic number Z ("Z-contrast"). The Al-rich shell is darker than pure GaAs – as shown in Figure 2b,c. However, at the specific conditions (detailed in the Methods and Supporting Information) MnAs NCs appear as bright objects.

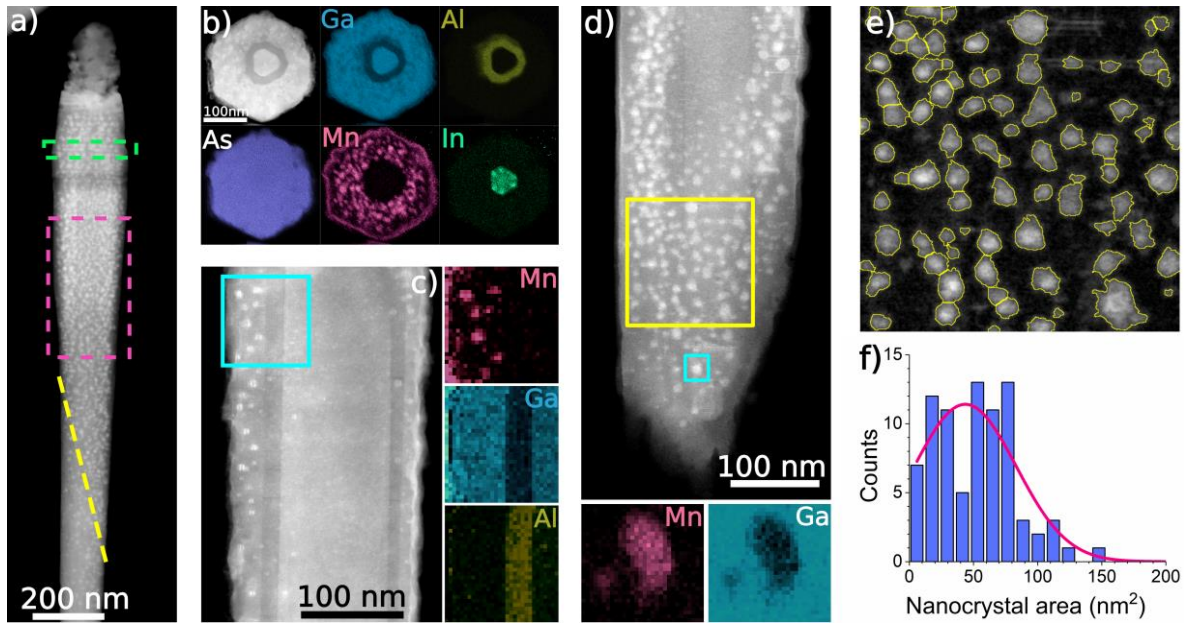

Figure 2. (a) Scanning transmission electron microscopy image of annealed NWs showing an individual NW with MnAs NCs visible as bright dots, (b) cross-section from the upper part of the NW perpendicular to the growth-axis (as schematically indicated by the green, dashed rectangle in (a)) with Energy-dispersive X-ray spectroscopy (EDS) maps of Ga, Al, As, Mn, In, of the whole area; (c) cross-section along the NW growth-axis (as schematically indicated by the pink, dashed rectangle in (a)) with EDS maps of Mn, Ga, acquired from the area marked by the blue square; (d) NW cross-section, at a small angle to the growth-axis (as schematically indicated by the yellow line in (a)) with EDS maps (Mn, Ga) acquired from the area marked by the blue square, proving that Mn appears in the Ga-depleted areas; (e) the section taken from the yellow square marked in (d) with subtracted background and outlined particle shapes (f) histogram of the particle cross-section areas distribution obtained from (e). Special conditions employed to obtain images in panel a-e are detailed in the **M**ethods.

A special specimen preparation is required to determine the distribution and shapes of individual MnAs NCs. To this end we use thin cross-section specimens cut perpendicularly and parallel to the growth axis as well as a small angle cut investigated in Figure 2b,c,d, respectively. This approach allows performing the image analysis to determine the NCs areas distribution which yields $44 \pm 6.0$ nm$^2$ as established from the Gaussian approximation presented in the panel (f), which corresponds to ~8 nm circle diameter. Such a low angle slice eliminates the other NCs from the field of view (from the sides and the rear part of the NWs) and enables the most accurate estimation of their lateral density in the shell, which amounts to $10^5$ NCs/μm$^3$ for the GaAs shell thickness of 50 nm.

The energy-dispersive X-ray spectroscopy (EDS) maps shown in Figure 2b,c,d prove the NW homogeneity and the expected elemental composition of the core and each shell. The EDS maps displayed in the panels (c) and (d) clearly confirm that the bright objects (NCs) correspond to the volumes containing Mn. One can observe in Figure 2d some Ga-signal present in the Mn-rich area, which comes from the GaAs matrix located under or over the MnAs NCs (common for the samples thicker than the NCs diameter). Moreover, one may clearly notice in Figure 2c that the (Ga,Al)As shell constitutes a diffusion barrier for Mn during annealing – Mn is detected at the level of noise < 0.2 at.% in the (Ga,Al)As shell and the core. However, as shown in Figure 2b, except of forming the MnAs NCs in the former (Ga,Mn)As shell and the outermost GaAs shell, an excess of Mn accumulates also at the NW surface. There is no direct comparison of the diffusion coefficients of Mn in GaAs and in AlAs available in the literature, but a compilation of theoretical and experimental values of the cohesive energy given in ref [32], clearly indicates that AlAs has a higher cohesion energy than GaAs, which may result in the lower diffusion coefficient of Mn in (Ga,Al)As in comparison to that in GaAs.

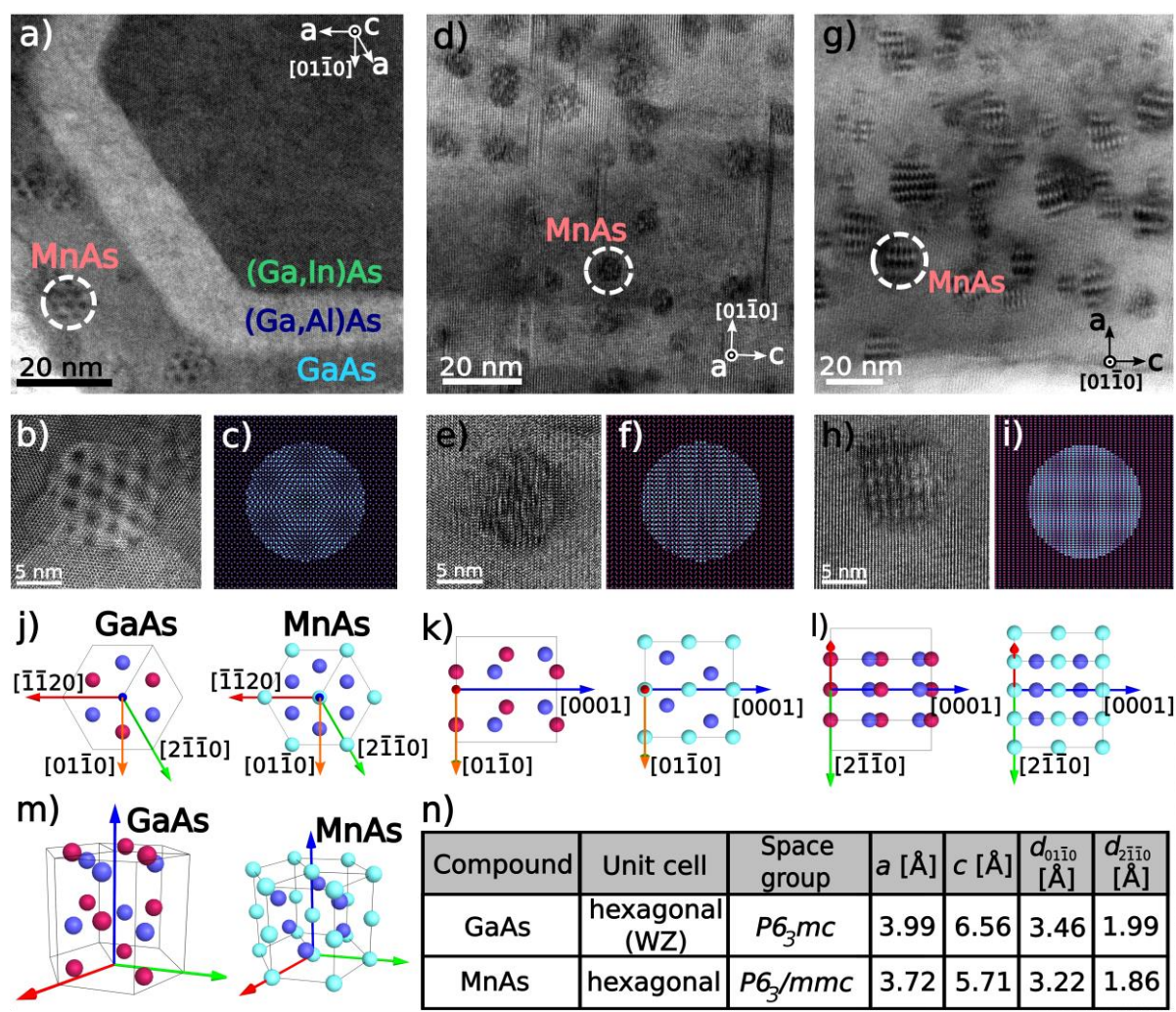

Figure 3. High resolution TEM images of FIB cross-sections of NWs with hexagonal MnAs NCs observed in three characteristic zone axes (a-b) [0001], (d-e) [2$\bar{1}\bar{1}$0] and (g-h) [01$\bar{1}$0]; compared with their model structures projections (c), (f), (i) respectively, created by overlapping (m) WZ-GaAs unit cell matrix and that of MnAs NC with hexagonal unit cell. The unit cells of both structures observed in (j) [0001] (k) [2$\bar{1}\bar{1}$0] and l) [01$\bar{1}$0] zone axes; (n) table with lattice parameters.[22,33]

Since TEM/STEM images are the planar projections of 3D structures, it is very challenging to prepare an ultra-thin specimen cut precisely through a small NC of interest without the surrounding WZ-GaAs matrix. Therefore the high resolution (HR) images with Moiré patterns resulting from the interference of these two similar structures are obtained, as exemplified in the panels (a) – (i) and explained in the panels (j) – (n) of Figure 3. Clearly, different patterns are observed, depending on the chosen zone-axis in the high resolution transmission electron microscopy (HRTEM) images. For each of the three investigated main zone axes, the images of single NCs are presented in the panels (b), (e), and (h) and compared to their structural ball model shown in the panels (c), (f), and (i), respectively.

The strain state is the essential feature of MnAs NCs embedded in the GaAs matrix since it directly influences their magnetic properties. The lattice mismatch:

$$M = \frac{d_{MnAs_b} - d_{GaAs_b}}{d_{GaAs_b}},$$

between the bulk (as the lower index $b$ indicates) hexagonal MnAs and the WZ-GaAs in reference to the WZ-GaAs is equal to -6.9% in the $[01\bar{1}0]$ direction and -13.0% in the $[0001]$ direction [from the values given in the panel (n) of Figure 3]. The formation of pseudomorphic (fully strained) MnAs NCs embedded in the WZ-GaAs lattice is precluded due to the extremely high strain energy at such a huge lattice mismatch. Similarly to the MnAs NCs embedded in the zinc-blende GaAs matrix, a certain degree of strain relaxation in MnAs NCs is expected here.[34] Hence, to enable this process, geometrical dislocations are necessary to appear at the interface and an additional atomic plane of MnAs should emerge every 2.27 nm in the $[0001]$ direction. As the result the MnAs NCs of diameters larger than 2.27 nm are expected to be relaxed, at least partially. To evaluate the MnAs NCs residual strain (RS) the geometric phase analysis (GPA) is applied. This method quantifies the displacements and strains in the crystalline structure employing digital signal processing algorithms based on the fast Fourier transform (FFT) of the HRTEM or HRSTEM images.[35]

The results of the GPA analysis are presented in Figure 4. We analyze the HRTEM image (Figure 4a) and HRSTEM image (Figure 4e) of the two NCs found in (a) the thicker and (e) the thinner cross-sections of the NW shells. The first one (Figure 4a) is completely surrounded by the GaAs matrix as explained by the 3D ball model shown in Figure 4i, whereas the second one (Figure 4e) represents an ultra-thin area crossing centrally through the NC, exposing its structure to the electron beam as visualized in Figure 4k. The ball projections shown in the panels (j) and (l) correspond well to the HR-images displayed in the panels a and e. The FFT analysis of the HR-images presented in Figure 4b,f confirms that the MnAs NCs are semi-coherent to the WZ-GaAs matrix (i.e. their unit cells, shown in Figure 3(j-l), have the same crystallographic orientations).

The components maps of the lattice distortion tensor $\beta$ – $\beta_{xx}$ and $\beta_{yy}$, common for the GaAs matrix and MnAs NC, are shown in Figure 4c,d,h. In our case the directions x and y correspond to the $[01\bar{1}0]$ and $[0001]$, respectively. It should be noticed that β is a result of the derivate of the displacement obtained from the lattice fringes common for MnAs and GaAs.[35,36] The $\beta_{xx}$ and $\beta_{yy}$ maps [the panels (c) and (d) in Figure 4] of the NC in the thicker cross-section (Figure 4a) reveal the crossing discontinuities inside the NC related to dislocations at the 3D NC/matrix interface. Regarding the NC in the thinner cross-section (Figure 4e), the $\beta_{xx}$ map (Figure 4h) reveals neither the presence of the crossing discontinuities (Figure 4c,e), nor the existence of the dislocations inside the NC. Characteristic spots around the NC perimeter seen in Figure 4h identify the positions of the misfit dislocations cores.

Since $\beta$ maps give an overall qualitative visualization of the projected lattice distortions, the average values of the lattice distortion inside NCs can be obtained by a direct measurement of the Bragg spots separation WZ-GaAs/MnAs in the FFT of HRSTEM images.[37] Also in this case the reciprocal space vectors g of the GaAs matrix lattice are used as a reference. The average value of $\overline{\beta}$ is obtained from:

$$\overline{\beta} = \frac{\frac{1}{g_{MnAs}} - \frac{1}{g_{GaAs}}}{\frac{1}{g_{GaAs}}}.$$

Finally, the average residual strain $\overline{\varepsilon}$ components inside the NC for specific direction are defined as the difference between $\overline{\beta}$ and M. For example, for the [0001] direction $\overline{\varepsilon}$ is given as:

$$\overline{\varepsilon}_{yy} = \overline{\beta}_{yy} - M_{[0001]} = \frac{c_{MnAs} - c_{MnAs_b}}{c_{MnAs_b}},$$
$$\overline{\varepsilon}_{xx} = \overline{\beta}_{xx} - M_{[01\bar{1}0]} = \frac{a_{MnAs} - a_{MnAs_b}}{a_{MnAs_b}}.$$

In the case of the NC shown in Figure 4a the $\overline{\beta}_{yy} = -12.5\%$ and $\overline{\beta}_{[01\bar{1}0]} = -5.7\%$ which give $\overline{\varepsilon}_{yy} = 0.5\%$ and $\overline{\varepsilon}_{xx} = 1.2\%$, respectively.

In this exemplary case, both residual strain components $\overline{\varepsilon}$ are positive, which means that the MnAs NCs are under anisotropic tensile strain. Similar analysis performed for dozens of such NCs gives generally positive magnitudes of $\overline{\varepsilon}_{yy}$ and $\overline{\varepsilon}_{xx}$, which range from -0.2% to 2% and 0.3% to 1.3%, respectively, i.e. $\overline{\varepsilon}_{xx}$ assumes always positives values.

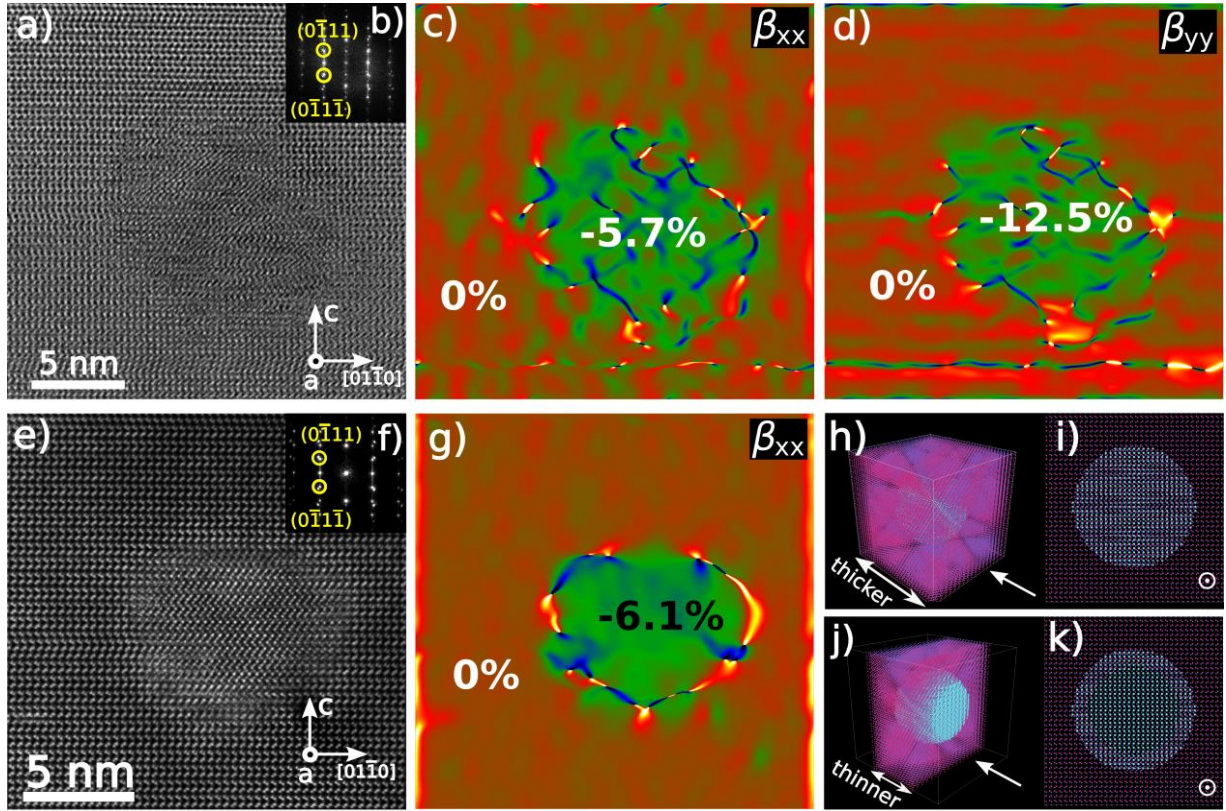

Figure 4. Geometric phase analysis (GPA) of (a-d) the MnAs NC completely surrounded by WZ-GaAs (a thicker specimen); (g-h) the MnAs NC centrally sectioned (a thinner cross-section). NW cross-section images acquired in [$2\bar{1}\bar{1}0$] zone axis in: (a) image-corrected HRTEM mode and (e) probe Cs-corrected HRSTEM mode; (b,f) Fast Fourier Transform of (a) and (e) respectively with two selected reflexes taken into account in the GPA lattice distortion evaluation; (c,d,h) The numerical strain components obtained from GPA defined with respect to WZ-GaAs lattice reference. (i,k) 3D ball model visualization of the MnAs NC in the WZ matrix of the thicker and thinner specimen, respectively; white arrows indicate the electron beam in TEM. (j,l) Ball model projection along the electron beam of (i) and (k), respectively, revealing differences in the single NCs images corresponding to HRTEM - (a) and (e) images.

In general, TEM investigations confirm that MnAs NCs embedded in the WZ-GaAs are under anisotropic tensile strain. Our HRTEM analysis shows that the NCs are stretched in all the directions, but the strongest effect occurs along the *c* axis. To our best knowledge no similar strain measurements for individual NCs embedded in GaAs are available in literature. The strain relaxation in the lower symmetry WZ-GaAs should be less effective than that in the cubic GaAs due to fewer equivalent primary slip systems (3 vs 12).[38] However, the mechanism of the lattice relaxation and misfit dislocation formation at the interface of WZ-GaAs matrix and MnAs NCs is extremely complex and needs additional theoretical and experimental investigations. Additionally, in the NWs analyzed here there are almost no voids associated with the MnAs NCs (with the few exceptions shown in the supporting information), which were ubiquitously evidenced in the previous reports concerning MnAs NCs embedded in the cubic GaAs [13-15] MnAs has a higher mass density than (Ga,Mn)As so the lack of voids is surprising. It can be explained by the proximity of free surfaces, which allow vacancies to out-diffuse towards the surface. On the other hand the lack of voids is the advantage of the NWs studied here, since it results in the increase of the strain state of the MnAs NCs hence it stabilizes the α-MnAs crystallographic phase up-to much higher temperatures. This directly leads to the shift of the Curie temperature associated with the α to β transition as documented below.

In the above analysis of strain exerted by WZ GaAs matrix on MnAs NCs we neglect the strain in the shells originating from $Ga_{0.92}In_{0.08}As$ core with the lattice parameter slightly higher than that of (Ga,Al)As-GaAs shells. Because of the similar volumes of the core and the shell materials in the NWs considered here, the longitudinal strain (along the NW axis) is equally distributed.[39] Hence for the $Ga_{0.92}In_{0.08}As$-GaAs core shell system with 0.5% lattice mismatch the longitudinal strain is about 0.25% in both components (with opposite sign for the core and the shells). This is over one order of magnitude less than the mismatch between WZ GaAs and MnAs considered above, thus it does not affect the GaAs-MnAs strain analysis in a significant way.

**Magnetic properties**. The magnetic properties of the as grown NWs are similar to those of the previously reported core-shell nanowire structures containing (Ga,Mn)As.[31] This underlines a very good command over the growth procedure and the achieved reproducibility of the investigated material. In short, the presence of Mn ions in the (Ga,Mn)As shell is responsible for the ferromagnetic (FM) coupling in this part of the NWs. In the case of the as grown material the coupling is effective only at the very low temperatures, as indicated in the panel (a) of Figure 5, and operates at mesoscopically short distances, as elaborated previously for similar NWs[30,31] and, generally seen in the low hole density (Ga,Mn)As.[30,40,41,42] Therefore, the magnetism of the thermally non-treated NWs assumes a heterogeneous form akin to that of nanomagnets, however, with this profound distinction that the whole Mn-rich shell is perfectly homogenous both chemically and structurally. None of the structural or element sensitive characterization tools can actually point the difference between these fragments that generate the magnetic flux and the rest of the sample.[43] So, as long as the growth conditions are not altered,[44] there can only be fluctuations in the local density of states that could end up in a patchwork of spontaneously magnetized volumes ($T_C > T_0$) containing a sufficiently high hole density to assure the FM spin-spin ordering within their phase coherence length. In this scenario the rest of the material with too low hole density to mediate the effective coupling remains nonmagnetic at the relevant temperature range ($T > T_0$),[40,45] where $T_0$ is the base temperature of an experimental setup.

A completely different picture is drawn by the annealed NWs. We start from a comparison of the temperature dependence of the magnetic moment measured at a moderate magnetic field of 1 kOe. The results shown in the panel (a) of Figure 5 are rescaled according to the area of the investigated ensembles of NWs. It is clearly seen, that both samples exert nearly the same magnetic flux at $T \cong T_0$, confirming the high lateral uniformity of the NWs density. On the other hand, as the magnetic response of the as grown NWs vanishes quickly above about 50 K, the annealed NWs stay magnetized in the whole temperature window available for the experiment, that is up to 400 K, and clearly above.

This finding corroborates our TEM results showing that the annealing triggers the crystallographic phase separation, upon which the Mn ions regroup and form the NCs, identified as MnAs, known to have the FM properties at room temperature. There is, however, more than just a switch to the granular nano-magnetic system. The magnetic signal clearly persists beyond the highest available temperature in our magnetometer (400 K) and even there it does not show a tendency to a critical fall down. This indicates $T_C$ in excess of 400 K, a value exceeding by far the magnitude of $T_C$ in MnAs as established previously[9,22,26] and evidenced here again upon the identically measured $m(T)$ dependence of a reference MnAs layer grown by us in the same MBE system as that used for NWs (the grey line in Figure 5a).

To approximate the magnitude of $T_C$ of our MnAs NCs more closely, we establish the values of the saturation moment $m_S$ from the magnetic isotherms $m(H)$ at high temperatures (detailed in the supplementary information in Figure S7). Then we plot $m_S$ as the function of $T$ (Figure 5b) together with the basic shape of the $T$-dependent Brillouin function adjusted to match the experimental results in the broadest possible range of $T$. The lack of a sharp decline of the experimental $m_S(T)$ at the critical region indicates the existence of a sizable distribution of $T_C$, which extends clearly above 400 K (~130 °C). The lower bound limit of this distribution has to be about 360 K, i.e. around the temperature where the experimental data start to deviate from the Brillouin function. Combining these two findings we can propose a scenario in which the MnAs NCs are strained so effectively by the NW environment that on increasing $T$ the FM α-phase is preserved far above 313 K. For some NCs the effect is so strong that they remain in the α-phase until or above 400 K, that is until the hexagonal γ phase sets in, importantly, maintaining the FM order all the way through. The experimental findings indicate therefore that the existing strain in the NWs bridges both hexagonal α- and γ- phases over the orthorhombic one. Most importantly this scenario indicates that the strain-induced stabilization of the hexagonal form of MnAs leads to the FM ground state extending to temperatures in which paramagnetic γ-MnAs emerges (in the bulk form). On the other hand it is possible that not all the NCs are strained so effectively, so they transform to the β-phase below 400 K and remain paramagnetic, so contributing marginally weak to $m_S$. This scenario is corroborated qualitatively by the first principle calculations presented in the next section.

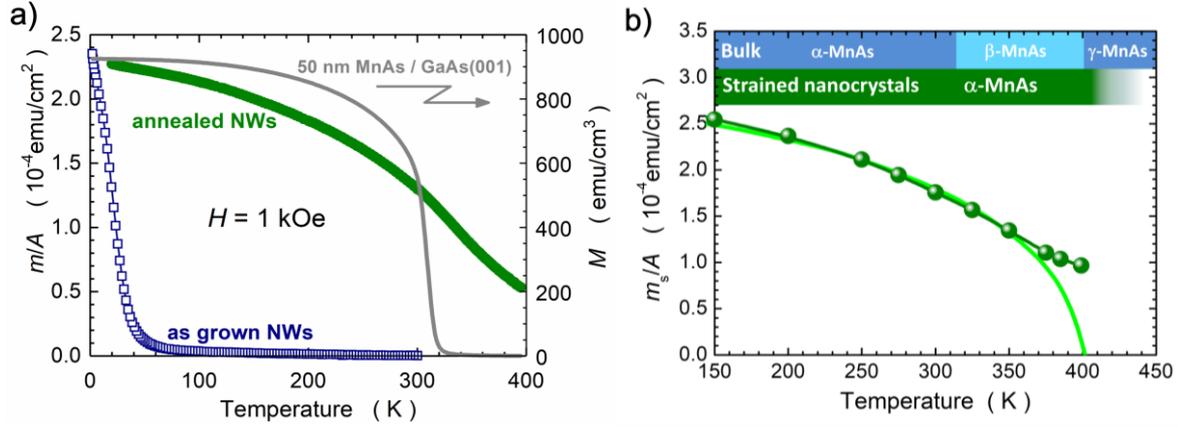

Figure 5. (a) Temperature dependence of the aerial density of the magnetic moment of the as grown (open symbols) and annealed (bullets) nanowires (NWs). Right Y axis refers to the magnetization of the uniform, 50 nm thick MnAs layer grown on GaAs(100), the grey line. All the measurements have been performed at magnetic field $H = 1$ kOe. (b) Saturated magnetic moment ($m_S$) of the annealed NWs (bullets) and the mean field Brillouin function (line) adjusted to the data to provide a lower-bound estimate of the Curie temperature of the NWs. The magnitudes of $m_S$ are established from magnetic isotherms presented in supplementary information Figure S7.

**First-principles investigation.** To understand the large increase of $T_C$ in the strained MnAs NCs embedded in the wurtzite GaAs matrix, we complement our analysis with the first-principles calculations. As mentioned above the MnAs NCs are exposed to the large tensile strains in all the directions but to a larger extent in the $c$ direction. On the other hand, there is a significant difference between the reported values of the elastic constants for α-MnAs in the $c$ direction ($c_{33} = 110$ GPa) with respect to those in the $xy$-plane ($c_{11} = 40$ GPa and $c_{12} = 8$ GPa).[46] As a consequence, in the strained MnAs NCs a larger strain relaxation is expected for $c$ than for $a$ lattice parameter. The elastic properties of solids are usually well described by first-principles calculations[47] (e.g. the $c_{33}$ value calculated by us for of bulk α-MnAs is 115 GPa), therefore, we believe that the $T_C$ changes in MnAs NCs as a consequence of stress will be also well captured.

The first-principles spin-polarized calculations are performed using a plane-wave basis, ultra-soft Vanderbilt-type pseudopotentials, and the Perdew-Burke-Ernzerhof exchange-correlation functional as implemented in the Quantum-ESPRESSO package.[48] The plane-wave cut-off is set to be 60 Ry (480 Ry for the electronic density) and a 8×8×8 $k$-point mesh for the Brillouin zone sampling is used. The calculated equilibrium values of the lattice constant for MnAs in the NiAs structure are $a = 3.717$ Å and $c = 5.545$ Å and agree well with the earlier theoretical reports.[49] The Curie temperature is calculated (Figure 6) within the mean-field approximation (MFA) to the effective classical Heisenberg Hamiltonian by using the following formula:

$$T_C = \frac{2}{3k_B}\sum_{j \neq 0} J_{0j} S_0 S_j \approx \frac{2 J_{NN} S^2}{3k_B} \frac{1}{12} = \frac{E_{AFM} - E_{FM}}{6k_B},$$

where $k_B$ is the Boltzmann constant, $J_{0j}$ is the exchange coupling constant of a given local spin $S_0$ with the $S_j$ spin ($S_0 = S_j = S$), $E_{AFM} - E_{FM}$ is the energy difference between antiferromagnetic (AFM) and FM configurations of magnetic moments (spins) localized on Mn ions and is used to evaluate the value of the exchange constant $J_{NN}$ between the nearest neighbouring magnetic moments.

The results of first-principles calculations are shown in Figure 6, where we plot $T_C$ as a function of $a$ and $c$ lattice parameters. At zero stress $T_C^{MFA} = 382\ K$, and it is by about 20% higher than the experimental bulk value of 313 K. Although the mean-field theory usually overestimates $T_C$ by 10-20%,[50,51,52] it is relatively much more precise in predicting the variation of $T_C$ with strain[51] or pressure.[52] Two observations are clear from the results of our calculations (Figure 6). First, $T_C$ weakly depends on $c$ (at least in the range of the relevant values of $c$) and second, $T_C$ considerably changes with $a$ in the vicinity of the zero stress (dashed line in Fig 6.). These results corroborate qualitatively the experimentally observed $T_C$ increase in the stretched MnAs NCs described in the previous sections. Our calculations unambiguously show the increasing trend of $T_C$ in tensely strained MnAs NCs, despite the fact that the exact values of $T_C$ are overestimated.

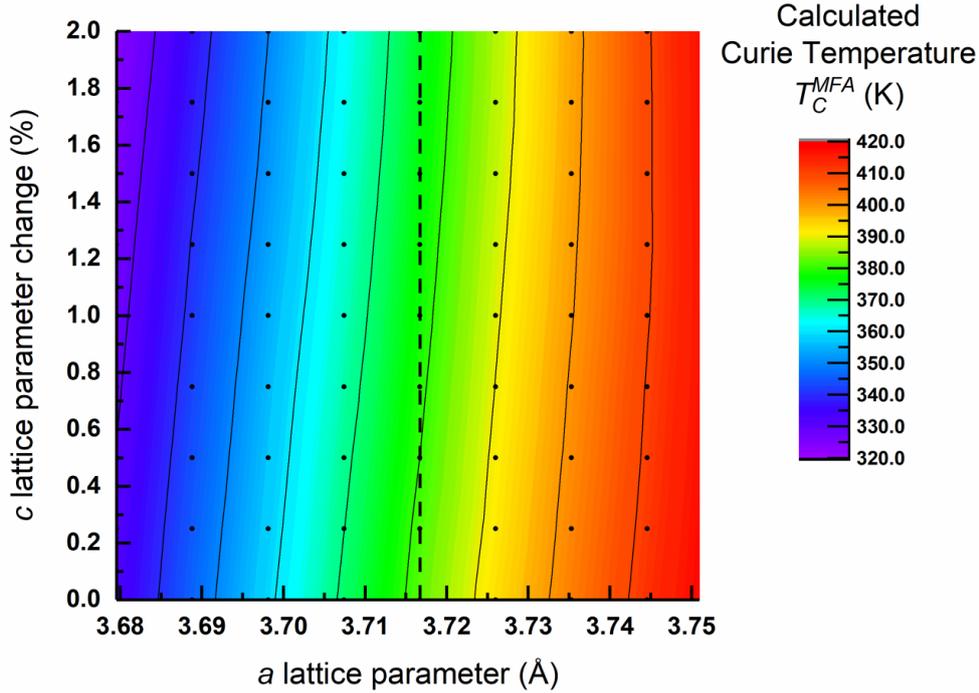

Figure 6. Curie temperature of MnAs in the NiAs structure calculated within the mean-field approximation (MFA) method as a function of $a$-lattice parameter and c-lattice parameter change. At zero stress $T_C^{MFA} = 382\ K$ and a = 3.717 Å. As long as a is close or above ~3.7 Å the structure of MnAs remains hexagonal.

CONCLUSIONS

In summary – we report on the huge, nearly 30% enhancement of the ferromagnetic phase transition temperature in strained MnAs nanocrystals embedded in the *wurtzite* GaAs matrix, as documented by the thorough magnetometry measurements. The effect is brought about by the stress exerted by the wurtzite surrounding, which stabilizes the MnAs nanocrystals in their hexagonal α-phase, the fact documented by the state of the art HRTEM and HRSTEM investigations. Our finding is further supported by the first-principles calculations pointing on the Curie temperature magnitude in the excess of 400 K in the tensely strained MnAs. The key result of our investigations is that the specific geometry of the WZ-GaAs:MnAs granular system prevents the usually occurring 2% MnAs lattice contraction associated with the ferromagnetic/hexagonal-to-paramagnetic/orthorhombic phase transition in the bulk MnAs. This mechanism opens wide avenues for the development of novel functional semiconductor-ferromagnetic hybrid nanostructures operating well above the room temperature.

METHODS

The NWs have been grown by molecular beam epitaxy (MBE) on GaAs(111)B substrates covered with (2 – 10 Å thick) pre-deposited gold layer. Gold has been deposited at room temperature, on epi-ready GaAs(111)B substrate surface, in a separate MBE system dedicated to metals. The substrates have been then transferred (in air) to the III-V MBE system equipped with the Mn source (Knudsen effusion cell). After introducing to the III-V MBE system and initial degassing of the substrates (at 200$^{o}$C) for several hours, the substrates have been transferred to the MBE growth chamber. The samples (Figure 1) have been first heated to the temperature of GaAs native oxide desorption (about 590 $^{o}$C), then the substrate temperature has been decreased to 490 – 560 $^{o}$C and the (Ga,In)As core NWs have been grown for about 3 hours (lower temperature was applied for (Ga,In)As core NWs). (Ga,Al)As shells have been deposited after completion of the axial growth at the substrate temperature decreased to 440 $^{o}$C. The substrate temperature has been further decreased to about 200 $^{o}$C for deposition of (Ga,Mn)As shell (Figure 1). For all the samples thin low temperature (LT) GaAs capping shell has been ultimately deposited (at the temperature of (Ga,Mn)As shell growth).

For the annealing procedures the as-grown samples, taken out of the UHV MBE system has been cleaved to pieces and reintroduced into the MBE system. Annealing has been performed in the MBE growth chamber with samples placed at the MBE substrate holders. For the precise control of the annealing temperatures the pieces of InAs(100) or GaAs(100) epi-ready substrates have been mounted on the same sample holder in parallel to the pieces of GaAs(111)B substrates with the NWs grown previously. Hence the absolute values of the annealing temperatures have been referenced to the temperature of the native oxides desorption from InAs(100) (lower annealing temperature of 450 ºC). Overall changes in NWs structure under annealing at these a selected temperature are shown in Figure 1.

Morphology and structure of the NWs have been investigated by the FEI Titan 80-300 transmission electron microscope (TEM) operating at 300 kV with spherical aberration corrector of objective lenses in the HRTEM mode. The STEM–HAADF images have been obtained for a different camera length of 70 mm to 280 mm, with converged semi angle of 9.5 mrad of the incident beam. The best contrast (brighter MnAs NCs embedded in darker WZ-GaAs matrix) has been obtained for the camera length of 230 mm and collection angles ranging from 25 mrad to 150 mrad where the signal registered by the HAADF detector originates from both high angle scattering and Bragg diffraction. Details are given in the supplementary information. The elemental composition determination was performed by EDS using EDAX 30mm$^2$ with collection angle 0.3 sr. Due to TEM investigations requirements the samples need special preparation procedures. In the first approach NWs have been transferred mechanically onto copper grids covered with holey carbon films (supported by Pacific Grid-Tech), enabling imaging the entire NWs. In the second approach, the cross-sections of resin-impregnated NWs have been cut using focused ion beam in Helios Nanolab 600 FIB as it is described in ref [53]. Most of the investigations have been performed with a double tilt holder enabling to orient the NWs to their easy zone axis. More detailed investigations (HRSTEM as well as STEM SuperX-EDS) have been performed with the probe Cs-corrected FEI TITAN$^3$ G2 60-300 microscope equipped with the ChemiSTEM system, operating at 300 kV, 29.5 deg. semi convergent angle of electron beam used for the atomic resolution STEM imaging. The lattice distortion measurements by Geometric Phase Analysis have been performed with a homemade software[35] using the probe Cs-corrected HRSTEM – HAADF images of FIB cross-sections of NWs.

For the magnetic studies the NWs have been separated at cryogenic temperatures from their substrates using an embedding film of PMMA [poly(methyl methacrylate), a popular e-beam lithography resist], as detailed in the supporting information file. Such a PMMA flake containing NWs has been attached to a customary cut long Si strips facilitating samples support[54] in the magnetometer chamber of a Quantum Design MPMS XL Superconducting Quantum Interference Device (SQUID) magnetometer operating in a wide temperature range $2 < T < 400$ K and in magnetic field $H$ up to 50 kOe. For all measurements $H$ has been applied perpendicularly to the wires, as this is the magnetically easy orientation for these NWs.[31] During the measurements the experimental code and data reduction methods specific for studies of minute magnetic signals have been strictly followed.[51]

ASSOCIATED CONTENT

**Supporting Information**

The Supporting Information containing detailed description of the TEM procedures and methods as well as sample preparation techniques and experimental procedures used for SQUID magnetometry measurements is available free of charge on the ACS Publications website.

AUTHOR INFORMATION

**Corresponding Authors**

*Email address: kaleta@ifpan.edu.pl

*Email address: kret@ifpan.edu.pl

*Email address: janusz.sadowski@lnu.se

ORCID
Anna Kaleta: 0000-0001-6034-6298
Slawomir Kret: 0000-0002-3532-5708
Janusz Sadowski: 0000-0002-9495-2648

**Author contributions**

J.S. designed the project and grew the MBE samples

A.K., S.K., M.S., J.S., K.G., S.B.K. contributed to writing the paper on equal basis

A.K., S.K., performed almost all TEM measurements

B.R. performed the probe Cs-corrected HRSTEM (Figure 4a,e) and Super-X EDS measurements (Figure 2b).

B.K. performed SEM images and prepared the TEM cross-section specimens

A.K., S.K., S.B.K. interpreted the TEM results

N.G.S. made and interpreted ab-initio calculations

K.G., M.S. performed SQUID magnetometry measurements and interpreted the results.

**Notes**

The authors declare no competing financial interest.

ACKNOWLEDGEMENTS

This work is supported in part by the National Science Centre, Poland under grants: 2014/13/B/ST3/04489, 2017/25/N/ST5/02942, 2016/21/B/ST5/03411, 2018/28/T/ST5/00503 and 2016/23/B/ST3/03575. J.S. acknowledges partial support by Carl Tryggers Stiftelse (Sweden) under the project CTS 16: 393. The authors thank Aloyzas Siusys from IP PAS for preparation of SQUID specimens, and Adam Kruk from AGH-UST for providing access to probe Cs-corrected (S)TEM at International Centre of Electron Microscopy for Materials Science.

# Supporting information

# Enhanced ferromagnetism in cylindrically confined MnAs nanocrystals embedded in wurtzite GaAs nanowire shells

*Anna Kaleta[1]\*, Slawomir Kret[1]\*, Katarzyna Gas[1], Boguslawa Kurowska[1], Serhii B. Kryvyi[1], Bogdan Rutkowski[2], Nevill Gonzalez Szwacki[3], Maciej Sawicki[1] and Janusz Sadowski[1,4]\**

1. **Optimization of scanning transmission electron microscope image contrast of MnAs nanocrystals embedded in WZ-GaAs.**

The difference of the visibility of MnAs nanocrystals (NCs) in the NW shell in the scanning transmission electron microscopy (STEM) image obtained from angular dark field detector (ADF) at two different camera lengths (CL) is obvious (Figure S1). At large camera lengths (more than 145 mm), a brighter contrast of the NCs in comparison to that of the matrix is observed. At CL of 145 mm there is no apparent contrast between the MnAs NCs and the GaAs matrix. With further CL decrease down-to 91 mm or less, some NCs start to appear again as dark areas surrounded by the brighter matrix (in other words the inversed contrast is observed), however, most of the NCs remain faintly visible in the STEM images. This is shown in Figure S1 (a), (b) – ovals 1 and 2. In the case of oval 3, additionally the void associated with the NC (indicated by arrow) is visible for both CLs. It should be emphasized that very few voids are detected. mainly in the upper NWs parts. The vacancies emerging during the transformation of (Ga,Mn)As dilute ferromagnetic semiconductor into GaAs:MnAs granular system have migrated towards the NWs surface and no voids have remained near the NCs. Only the individual voids appear due to the condensation of vacancies at stacking faults, as shown by arrows in Figure S1 (a), (b).

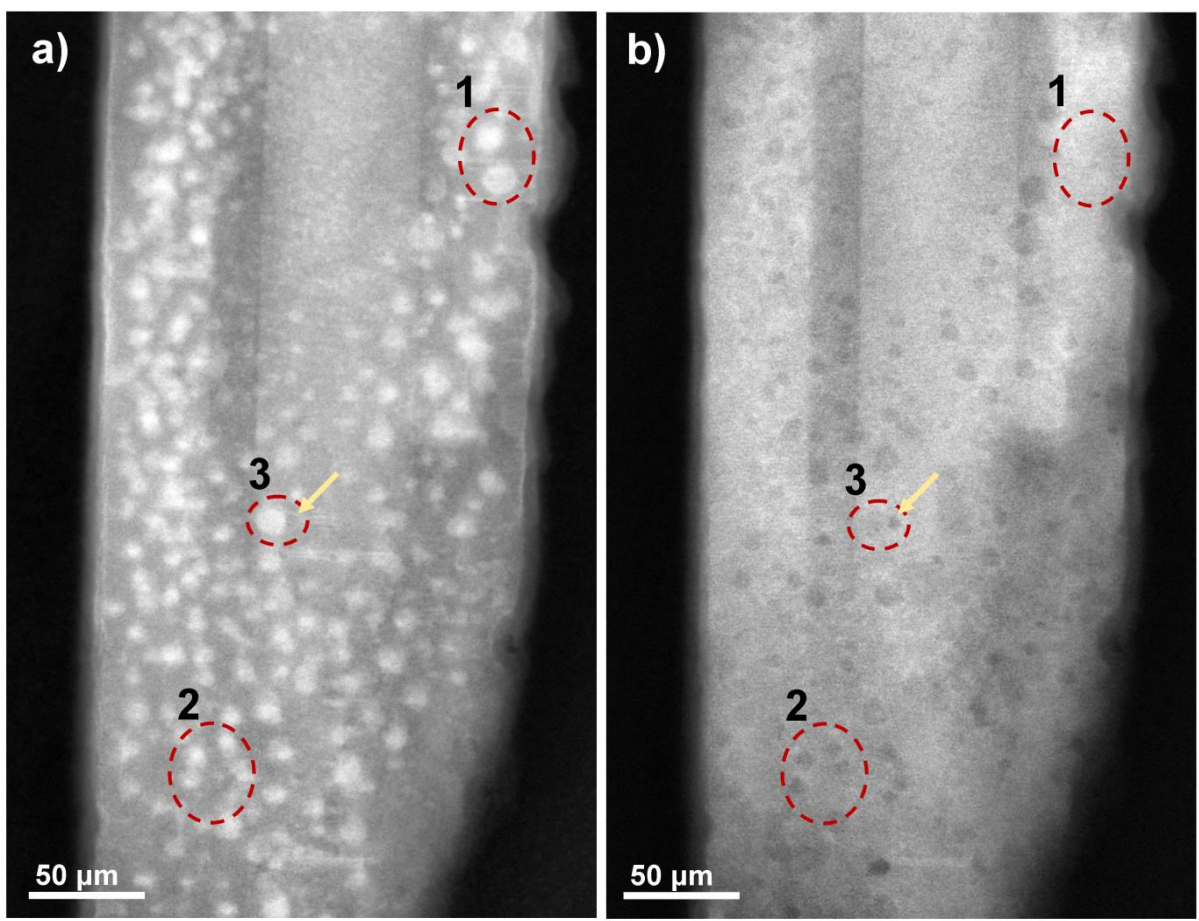

Figure **S1.** Scanning transmission electron microscopy (STEM) annular dark field (ADF) images obtained from oblique cross-section of a nanowire at different camera lengths: (a) 230 mm camera length and (b) 73 mm camera length. The dashed ovals indicate three characteristic places: 1 – the NCs vanish at short CLs, 2 – the NCs start to appear as dark areas surrounded by the brighter matrix, 3 – the void (indicated by arrow) associated with the NC is visible for both CLs.

    This contrast dependence on CL cannot be linked directly to the so called "Z-contrast" due to the contributions from diffraction components (Bragg spots) to the resulting STEM image. The classical "Z-contrast" can be observed only for amorphous materials and individual atoms. In the case of crystals the situation is more complex due to diffraction of the electron beam caused by periodic features. To reduce the influence of the diffraction contrast in the case of crystalline samples the low CLs are used to record the STEM images and to obtain preferential mass-thickness contrast. STEM images recorded at high CL are dominated by the diffraction contrast. In our case the experimental images obtained for short CL present very weak contrast between WZ-GaAs and MnAs. In contrary, for CL larger than 230 mm the MnAs NCs appear brighter than the GaAs matrix. This is due to the

strong Bragg spots contribution to the intensity acquired on ADF detector because of the smaller lattice parameter of MnAs (more Bragg spots reach detector in comparison to the diffraction from the GaAs matrix). Similar effect has been observed before for MnAs NCs embedded in the cubic GaAs matrix.[1]

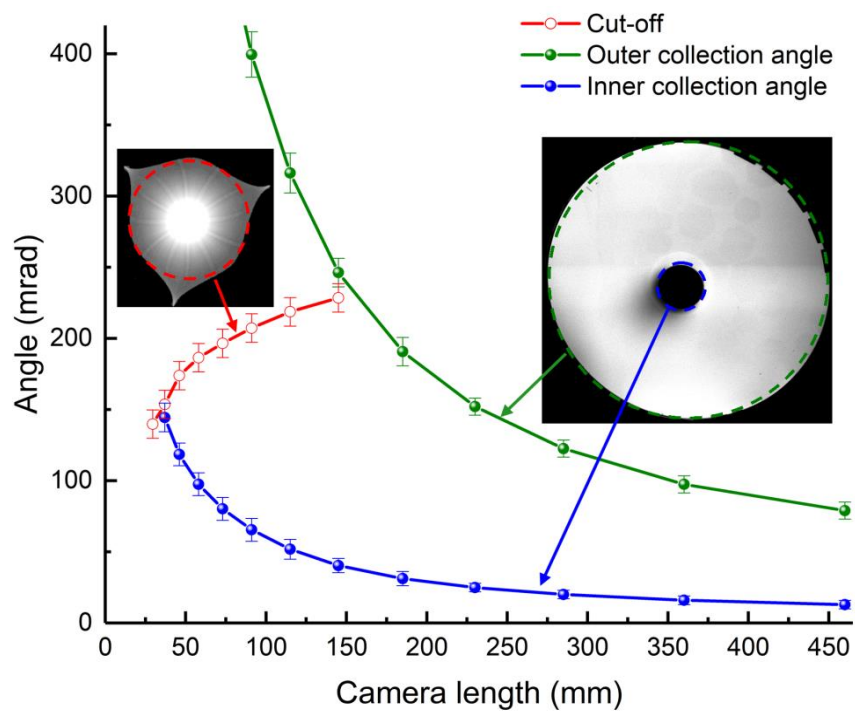

Figure **S2**. The dependence of ADF collection angles on the camera length. The insets show the shape of cut-off function for uncorrected STEM mode at 300 kV and the scan of the ADF detector (Fischione model 3000) together with the high angular cut-off of diffraction pattern with characteristic tree-fold features. Dashed circles represent cut-off used in the STEM simulations.

To optimize the visibility of NCs in the WZ-GaAs matrix the series of STEM image simulations were performed. Firstly, it is necessary to determine the collection angles for each CL of Titan 80-300 microscope equipped with the image corrector, used in our research,. The dependence of the collection angles on the camera lengths is presented in Figure S2. The upper (green) and lower (blue) curves correspond to the outer and inner ADF detector collection angles, respectively. Their ratio is limited by the active scintillator area and does not depend on the CL. This ratio is determined to be equal to 6.1±0.2. The inner angle is measured using a beam deflection, the procedure described in details in ref 2. The outer angle can be determined as the inner angle value multiplied by the measured physical ratio (6.1 here). However, the impact of the cut-off angle of the microscope optics on scattered electrons is significant for CLs below 145 mm.[3] This happens when the transferred diffraction has lower angular range than the outer ADF detector at small CLs. Because of that, not the entire active area of ADF detector is illuminated. Moreover, almost all the diffraction patterns reach the hole of the detector at 29.5 mm CL. Therefore, the usable outer ADF detector collection angles are restricted by a smaller cut-off value or the scintillator area limitation. However, the accurate determination of the cut-off angles is difficult due to the irregular shape and nonlinear scale of the diffraction patterns at the edge.[4] Taking into account these effects, the real clipping angles may even be somehow larger than those shown in Figure S2.

The STEM image simulation was carried out using the QSTEM software [5] to verify the contrast as a function of the CL. The chosen simulation parameters correspond to the experimental conditions applied during image acquisition presented in the main paper: 300 kV acceleration voltage, 1 eV energy spread of the electron beam. 9.5 mrad convergence angle, spherical aberration coefficient Cs for not aberration corrected Titan with Super-Twin probe forming lenses of 1.2 mm and Scherzer defocus. The In simulations, the maximum scattering angle is limited to 262.5 mrad. The chosen detector scattering angles range correspond to the values presented in Figure S2. The simulation is carried out without taking into account the thermal diffusion scattering.

To evaluate the dependence of the contrast on the CL, we consider a simplified model based on MnAs and GaAs single layers. The main idea is to compare the scattered intensities for the comparative thicknesses of the individual layers of MnAs and GaAs. This approximation applies to sectioned NWs, were the MnAs NCs sizes are comparable to the GaAs matrix thicknesses. In the model we use GaAs and MnAs layers of 100 unit cells thicknesses in $<11\bar{2}0>$ crystallographic direction. The simulated area is decreased to 25*25 Å for GaAs and MnAs layers with 0.25 Å pixel size.

The contrast is calculated based on the following equation for simulated and experimental images for all thicknesses:

$$C = \frac{I_{MnAs} - I_{GaAs}}{I_{GaAs}} \times 100\% \quad (1)$$

where $I_{MnAs}$ and $I_{GaAs}$ are the intensities of MnAs and GaAs calculated as the sum of the same areas of simulated and experimental STEM images.

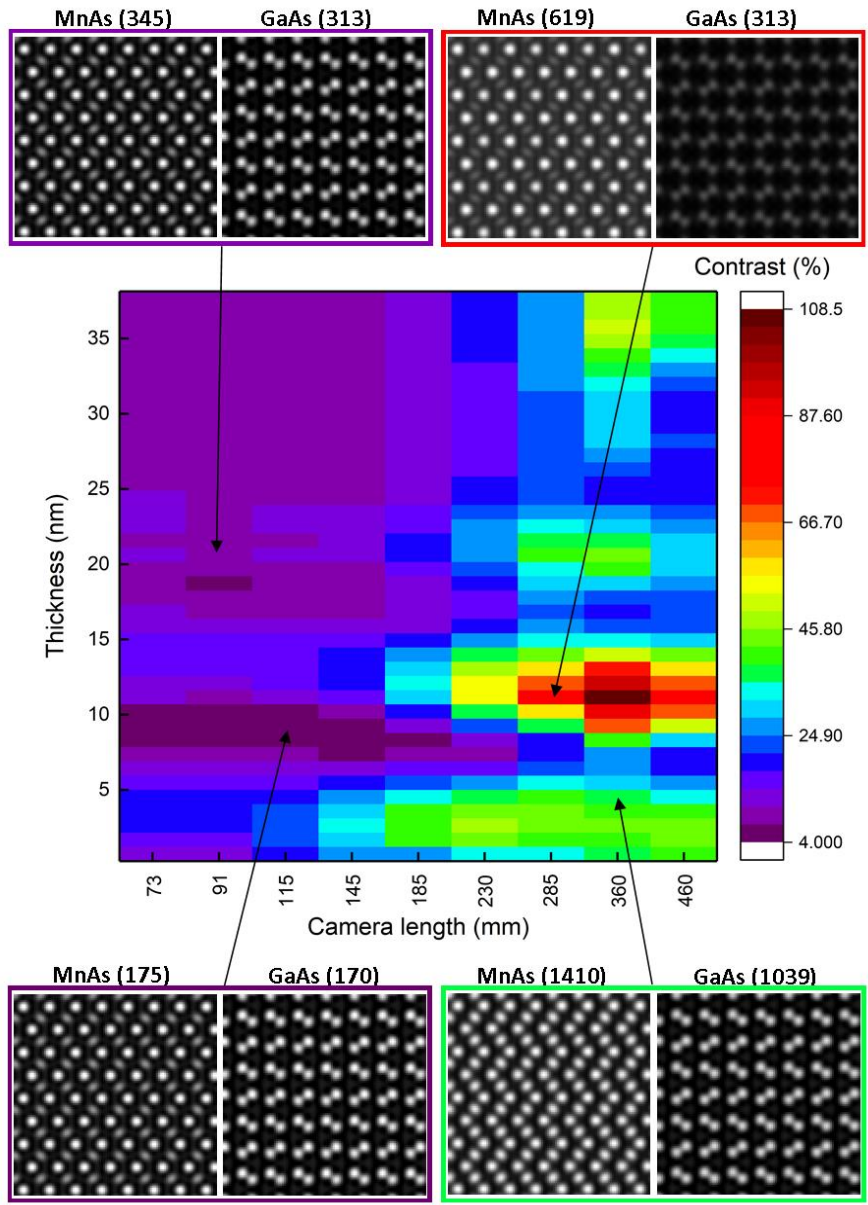

Figure **S3**. The contrast map calculated for the same thicknesses of MnAs and GaAs layers up to 37 nm (y-axis) and for CL range 73-460 mm (x-axis). For selected pixels (indicated by the arrows in the map) the pair of simulated STEM images of MnAs and GaAs layers in the same intensity scale is given. The numbers in parentheses refer to the sum of intensities of the simulated STEM image.

Figure S3 shows the contrast map calculated for simulated STEM images. Each pixel represents contrast value calculated using equation (1). For all the thicknesses at CLs less than 145 mm a relatively weak change in contrast is obtained. For larger CLs the contrast change is much more pronounced. This is due to the channeling effect. More precisely, the extinction lengths are slightly different for GaAs and MnAs (Figure S4), which causes mutual shift of the intensity maxima and significantly affects the contrast.

Similar effect of MnAs and GaAs channeling, is observed for different atomic Mn, As and Ga columns in the simulated data. Due to the redistribution of the intensity along the atomic columns, it is manifested as vanishing of some columns in the STEM image. We notice the enhancement of this effect with changed/increased CL in the simulated images. The effect is more pronounced for MnAs with more significant difference between the atomic numbers of Mn (25) and As (33) - compare simulation patters pairs in Figure S3.

These simplified simulations do not entirely agree with the experimental observation. The experimental contrast is calculated similarly as a simulated one. The obtained contrast value is equal to -4±3% for 73 mm CL, about 0±3% for 91-145 mm CLs, 13.4±4% for 230 mm CL and 15.4±4% for 285 mm CL. As can be seen in Figure S3, there are no pixels with a negative value. Nevertheless, the enhancement of the contrast with the CL increase, confirmed by the experiment, is clearly manifested here. The smallest contrast value, displayed in Figure S3 is about 4% which is close to the experimental results at small CL but still does not reach a negative value.

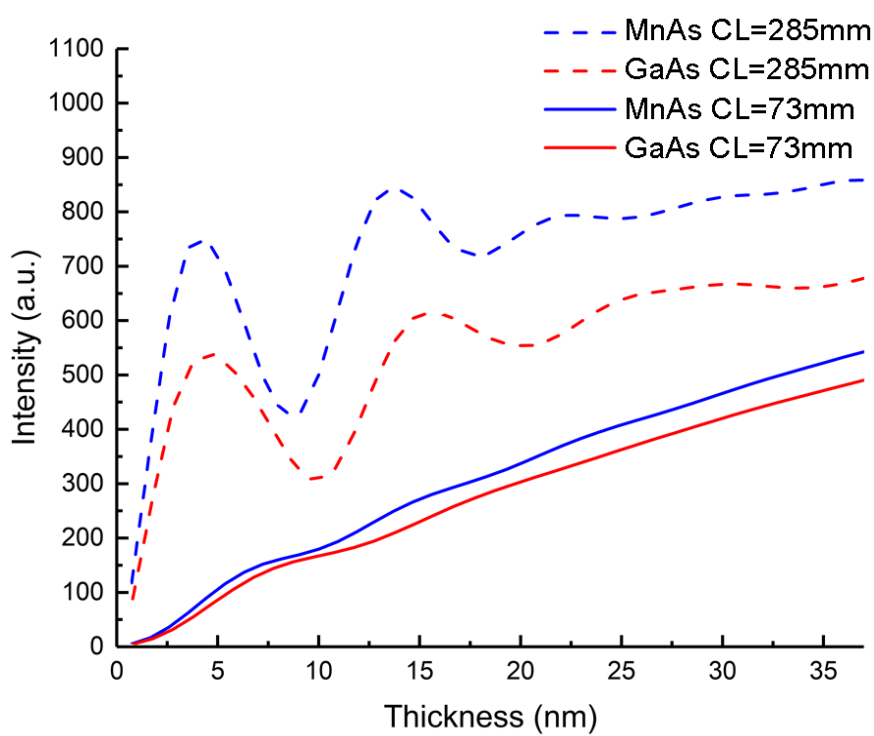

Figure **S4**. The average intensities of the simulated STEM images of MnAs and GaAs crystals as a function of thicknesses for 73 and 285 mm CLs. The effect of channeling is clearly visible for 285 mm CL.

To clarify this, more realistic simulation based on the model of spherical 8 nm hexagonal MnAs NC embedded in the WZ GaAs matrix is performed. The resulting contrast map is presented in Figure S5. The areas from MnAs NC and GaAs matrix indicated in Figure S5b by orange frames, are used to calculate the contrast values. The homogeneous horizontal band visible at the contrast map up to 2 nm (Figure S5a), with close to zero contrast, indicates that the simulation still does not reach MnAs NC. The embedded NC emerges at a 2 nm depth.

In the contrast map shown in Figure S5a, two characteristic regions with negative (73-115 mm) and positive (185-460 mm) CL are clearly visible confirming qualitatively the experiment observations. It should be noted, that the most pronounced positive and negative contrast occur at different depths. This occurs due to the channeling effect as described above. and additionally influences the radial contrast variation visible in Figure S5.

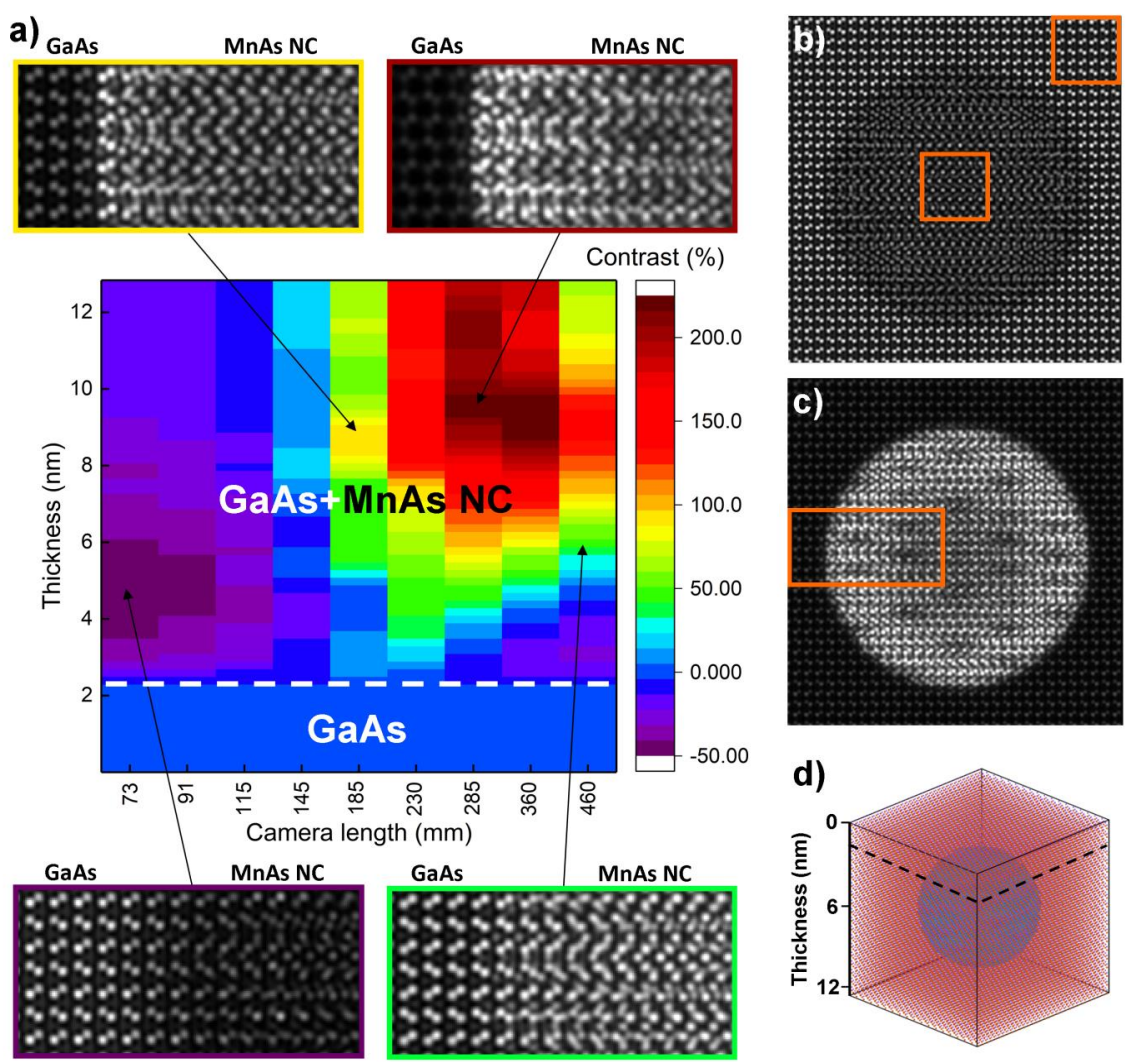

Figure **S5**. (a) The contrast map calculated for the 8 nm-thick MnAs nanocrystal embedded in the GaAs matrix for the CL range 73-460 mm (x-axis) and cumulated depth (y-axis). For the selected pixels (indicated by arrows) in the map the small areas of simulated STEM images are given. (b) Simulated STEM image for 73 mm CL and 12.7 nm thickness, frames indicate areas for calculation contrast map. (c) Simulated STEM image for 285 mm CL and 12.7 nm thickness. The indicated area represents areas of simulations given in (a) panel. (d) 3D view of the model.

To summarize this contrast analysis – the best visibility and signal-to-noise ratio of MnAs NC embedded in WZ GaAs matrix can be obtained using ADF Fischione 3000 detector and CL ranging from 230 to 360 mm. In these conditions the image contrast is dominated by contribution of MnAs Bragg spots to the image intensity without WZ-GaAs Brag spots contribution, related to the difference in the lattice parameters of WZ-GaAs and MnAs crystals.

## 2. Magnetic measurements

### 2.1 Sample preparation

Except of the NWs there are two other sources of the magnetic signal present on the typical "out-of-the-growth-chamber" specimens: the bulky semiconductor substrate, and the material that has non-intentionally grown in-between the NWs. The first component, the diamagnetic response of the substrate, GaAs in this case, does not pose any thread to the fidelity of the current studies. Given the sufficiently substantial magnitude of the magnetic flux exerted by Mn-containing NWs – a linear in magnetic field $H$ and temperature $T$ independent diamagnetic component of GaAs can be established independently and subtracted from the final results, rescaled properly to match the weight of the NWs-carrying sample and its physical dimensions.[6] The second component constitutes a far greater experimental problem , since its magnetic properties are *a priori* completely unknown. This is mainly because very little is usually known about the stoichiometry of the compound which has been deposited in-between the NWs during the growth of NWs cores and their subsequent overgrowths. In addition, the shadowing effect plays its role too, so neither the chemical composition nor the crystallographic structure are established, particularly that these properties are usually out of the scope of the research. In the case of the present studies the presence of magnetic species there makes the contribution of these layers to the total magnetic response significant.

In order to completely eliminate both spurious contributions to the magnetic signal we adopt the same method used in our previous studies of NWs containing (Ga,Mn)As,[7,8] proposed originally in ref 9. The method relies on a large difference of the thermal contraction between the semiconductor substrate and typical poly-resin. We use poly(methyl methacrylate) (PMMA, the commonly used e-beam resist), which coats the NWs-containing surface to a thickness of a small fraction of a millimeter, and then subsequently slowly cool the whole sandwich down to the cryogenic temperatures. There, the PMMA layer peels off taking all the NWs with itself. Such a NW containing PMMA flake is sufficiently robust and constitutes a convenient specimen for magnetic studies. The main advantage of this method is that the layer contracting PMMA reaps only the NWs, so both the whole substrate and the parasitic layer grown in-between the NWs is left behind. The effectiveness of this process is exemplified in Figure S6, where we present birds eye views over a few areas of the XYZ sample after a large PMMA flake peeled off with the NWs from the substrate. Equally important for magnetic studies, this method also preserves the mutual orientation of the NWs, what can facilitate magnetic anisotropy studies. On the other hand, the employment of PMMA introduces another magnetic response, so it is of a paramount importance to experimentally quantify its magnitude. The magnetic tests of the PMMA used for the separation of the NWs in this study indicate only a weak, linear in $H$ diamagnetic response at elevated temperatures relevant here. It gets only marginally modified by a small paramagnetic components at the lowest temperatures, $T < 10$ K, which can be disregarded in these predominantly high temperature studies. The flakes obtained here of both as grown and annealed ensembles of NWs prove sufficiently robust to withstand the whole suite of magnetic measurements without any additional supporting carriers or casings.

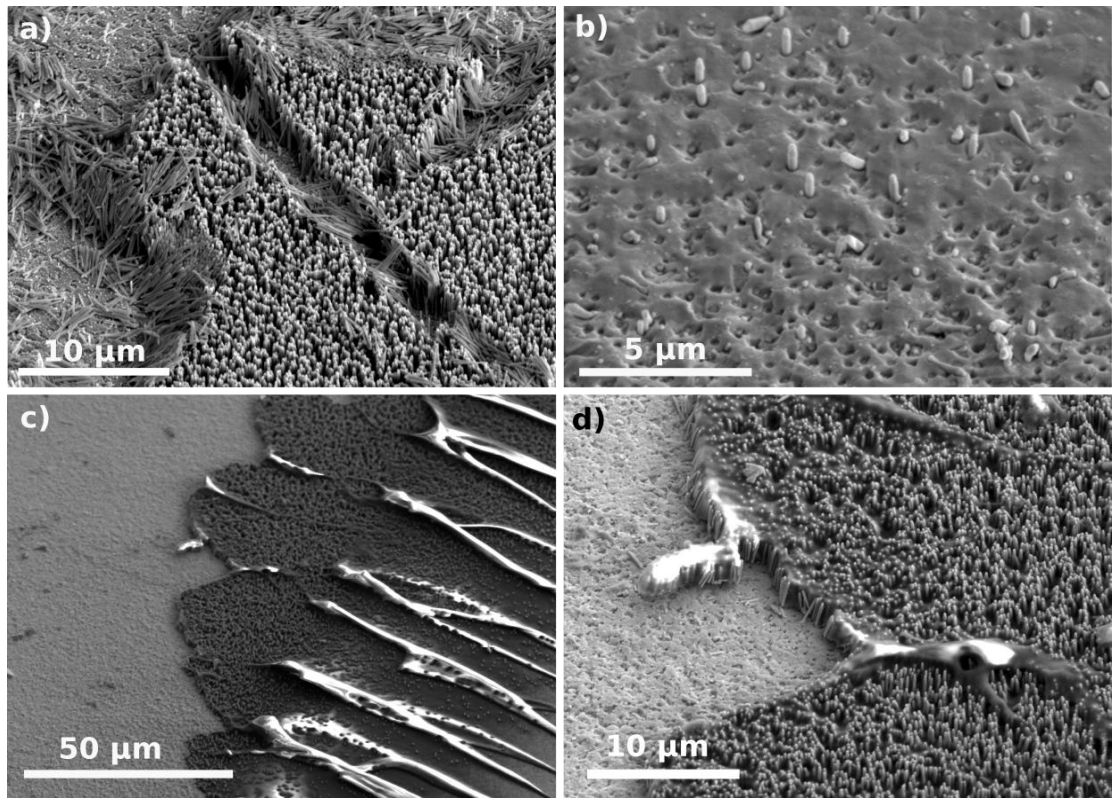

Figure S6. Tilted scanning electron microscope images of some fragments of the surface of the NWs sample after peeling off the PMMA. (a) A mechanically disturbed edge part of the sample which has not been covered with PMMA. The mechanically floored or removed NWs unravel the state of the surface after complete removal of the NWs. (b) A close up view on that part of sample form which NWs are lifted by the PMMA. A clearing-like appearance of the surface indicates a successful transfer of the NWs into the PMMA flake. (c-d) Images taken near another edge of the substrate, where the thickness of the PPMA has been too small to break the bonds between the NWs and the substrate (right hand sides of both images).

## 2.2 Determination of the saturation magnetic moment at high temperatures

We establish the values of the saturation moment $m_S$ of the MnAs NCs in the NWs from magnetic isotherms $m(H)$ measured at high temperatures. The results are presented in Figure S7, where the original magnetic data sets are already corrected for the diamagnetic signal of the PMMA material (flake) embedding the NWs. The diamagnetic contribution of the PMMA has been obtained from the high temperature $m(H)$ measured for another flake containing the non-annealed NWs and rescaling it according to the weights of both flakes. In this approach we take advantage that the magnetic signal of the paramagnetic Mn ions in the as grown (Ga,Mn)As shells is vanishingly small in comparison to the signal of the PMMA flake at elevated temperatures.

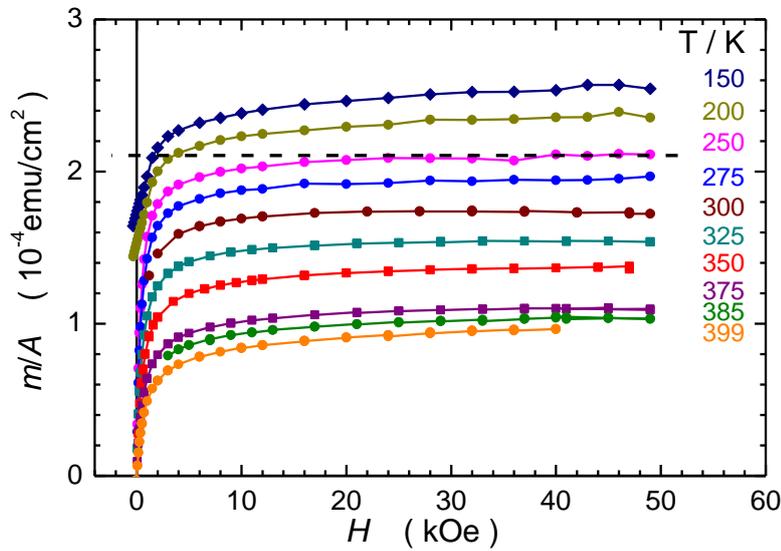

Figure S7. Aerial density of the magnetic isotherms $m(H)$ of the annealed NWs. The results are corrected for the diamagnetic signal of the PMMA flake embedding the NWs. The dashed black line exemplifies the establishment of the relevant saturation magnetic moments at 250 K.